\documentclass[]{iau}

\usepackage{amsmath}
\usepackage{graphicx}
\usepackage{multirow}

\def\aap{A\&A}

\begin{document}

\lefttitle{Guiglion et al.}
\righttitle{Realising the potential of large spectroscopic surveys with machine-learning}

\jnlPage{1}{7}
\jnlDoiYr{2025}
\doival{10.1017/xxxxx}

\aopheadtitle{Proceedings IAU Symposium}
\editors{}

\title{Realising the potential of large
spectroscopic surveys with machine-learning}

\author{G. Guiglion$^{1\star,2,3}$}
\affiliation{
$^{1}$Zentrum f\"ur Astronomie der Universit\"at Heidelberg, Landessternwarte, K\"onigstuhl 12, 69117 Heidelberg, Germany\\
$^{2}$Max-Planck Institut f\"{u}r Astronomie, K\"{o}nigstuhl 17, 69117 Heidelberg, Germany \\
$^{3}$Leibniz-Institut f\"ur Astrophysik Potsdam (AIP), Potsdam, Germany\\
$\star$ gguiglion@aip.de}

\begin{abstract}
Machine-learning is playing an increasing role in helping the astronomical community to face data analysis challenges, in particular in the field of Galactic Archaeology and large scale spectroscopic surveys. We present recent developments in the field of convolutional neural-networks (CNNs) for stellar abundances in the context of the Galactic spectroscopic surveys \emph{Gaia}-ESO, and \emph{Gaia}-RVS. Especially, by combining the full \emph{Gaia} data product, we manage to characterize for the first time the [$\alpha$/M] vs. [M/H] bimodality in the Galactic disc with \emph{Gaia}-RVS spectra at low-S/N. This work is highly relevant for the next generation of large scale surveys such as MSE, 4MOST, and WST.
\end{abstract}


\maketitle

% ------------------------------------------------------------
\section{Introduction}

Precise stellar chemical abundances are crucial for constraining the formation and evolution of the Milky Way and its neighbouring galaxies, as they allow stars to be used as fossil
records of past star formation events and enable the disentangling of stellar populations or the tracing of accreted stars
and stellar streams (e.g. \citealt{Matteucci2021}). The community has developed within the last two decades specific spectroscopic surveys to study in great details the Milky Way, for instance Gaia-ESO \citep{Gilmore2022} and APOGEE \citep{Majewski2017}, targetting several $10^5$ stars in the main MW components, with the ultimate goal of providing the community precise and accurate chemical abundance diagnostics for Galactic Archaeology studies. Standard spectroscopy is the unique way to measure detailed and precise chemical abundances and is well suited for rather small sample of spectra to analyse (up to $10^5$ stars).

% ------------------------------------------------------------
\section{Convolutional Neural-Networks (CNNs) for stellar labels determination}

New, extremely large spectral surveys have pushed standard
spectroscopic techniques to their limit, requiring a fundamental shift in the spectral analysis methods. Machine learning
(ML) methods have been used to propagate the knowledge of
standard spectroscopy to large spectroscopic datasets. The fundamental idea is to build a set of reference stars (training sample) with atmospheric parameters and chemical abundances (stellar labels) determined using standard spectroscopy. A ML model
is then built between stellar spectra and stellar labels and propagated to an external set of spectra. Such a method is powerful
because it allows for the simultaneous derivation of many stellar labels for millions of spectra, typically in several minutes. Several types of ML algorithms have recently been employed in the Galactic Archaeology community in order to parameterise stellar spectra. For instance, Convolutional neural
networks (CNNs), which build a model between spectra and
labels. CNNs are well known for being sensitive to spectral features and learning from such features, as well as being less sensitive to radial velocity shifts in the spectra than simple artificial neural networks (see for instance \citealt{Nepal2023}). The CNN allows to build of a high-dimensional non-linear function that translates spectra (plus extra data) to stellar labels (see for instance \citealt{Leung2019, Fabbro2018, Guiglion2020} for extensive details on CNNs).\\

Recent applications of CNNs on \emph{Gaia}-ESO survey stellar spectra showed that CNNs are able to learn from relevant spectral features for a given label, for instance Li feature for Li abundance measurements (see Figure 8 of \citealt{Nepal2023}). By computing gradients, ie. derivatives of a given label as a function of wavelength, one can visually inspect where the network learns. Hence, it is easy to check if the network learns from relevant features or from astrophysical correlation (see for instance Figure 11 \& 12 from \citealt{Ambrosch2023}).

% ------------------------------------------------------------
\section{The [$\alpha$/M] vs. [M/H] bimodality as traced by \emph{Gaia}-RVS}

Out of the $10^6$ RVS spectra provided by the \emph{Gaia} consortium, about 40\% have 15$<$S/N$<$25, for which no parameters or abundances with good flags are provided by the GSP-Spec pipeline. In \citet{Guiglion2024}, we aimed at combining the full \emph{Gaia} data product together with a hybrid-CNN to determine supercharged atmospheric parameters and chemical abundances from the DR3 \emph{Gaia}-RVS spectra, and leverage the low-S/N ratio RVS spectra. The data used in the \citet{Guiglion2024} consists of \emph{Gaia} DR3 RVS spectra \citep{GaiaCollaboration2023}, together with \emph{Gaia}
DR3 photometry (phot\_g\_mean\_mag G; phot\_bp\_mean\_mag
G\_BP; and phot\_rp\_mean\_mag G\_RP magnitudes;), parallaxes \citep{Lindegren2021}, and XP coefficients \citep{DeAngeli2023}. The labels of the training sample are
from APOGEE DR17 \citep{Abdurrouf2022}.\\

\begin{figure}
\centering
\includegraphics[width=0.99\linewidth]{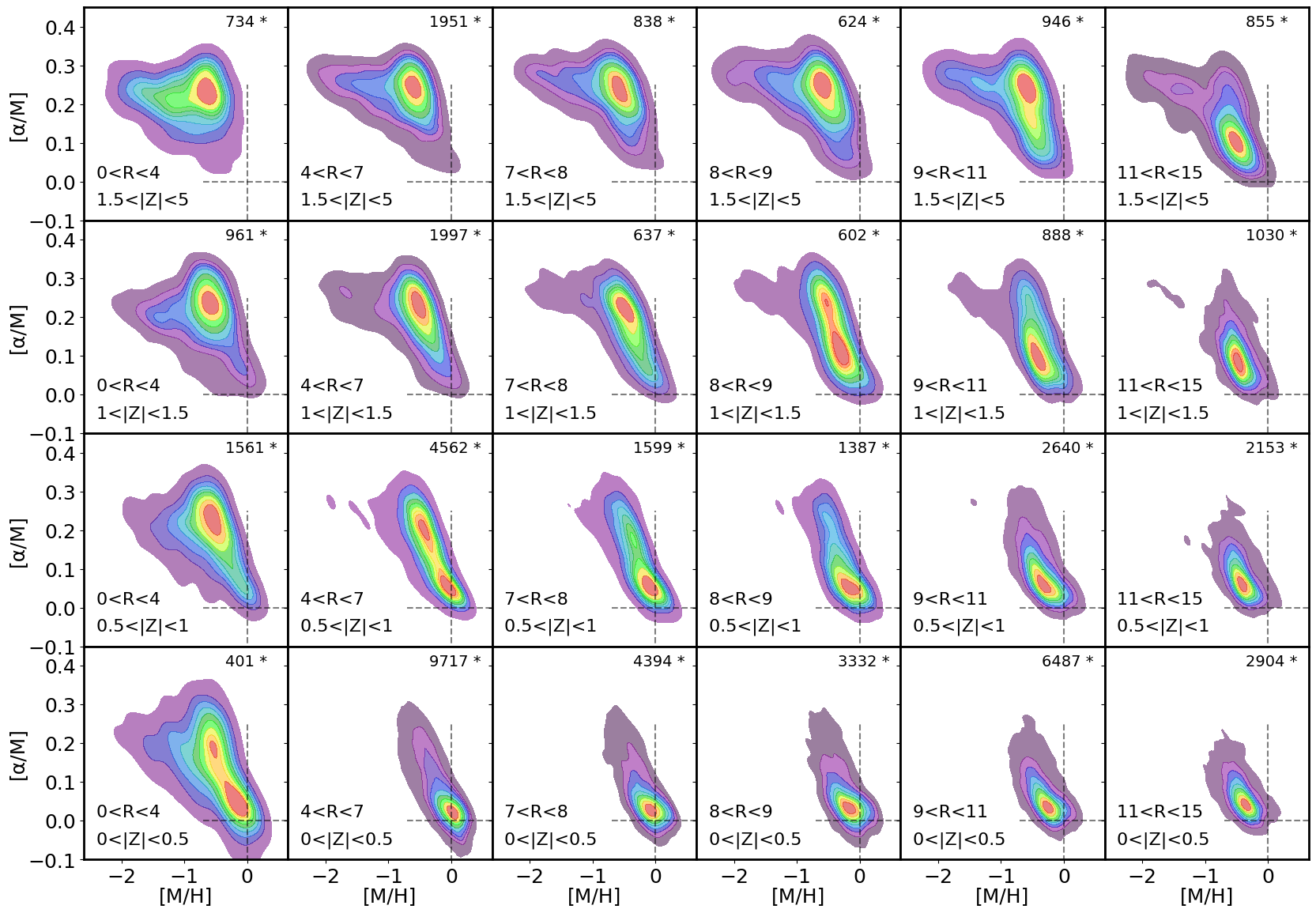}
\caption{Two-dimensional histograms and contours of [$\alpha$/M] vs. [M/H] in 53\,200 giants of the \emph{Gaia}-RVS observed sample from \citet{Guiglion2024} with 15$<$S/N$<$25 and log(g)$<$2.2 within the training set limits. Stars are plotted in kiloparsec bins of galactocentric radius (R) and height above the Galactic plane (Z).}
\label{fig:li}
\end{figure}

In \figurename~\ref{fig:li}, we show the [$\alpha$/M] vs. [M/H] plane decomposed into bins of R \& Z for $53\,200$ \emph{Gaia}-RVS stars with 15$<$S/N$<$25 and log(g)$<$2.2 from \citet{Guiglion2024}. The bimodality is for instance clearly seen in the bulge region with our CNN abundances, confirming previous results based on APOGEE DR14 and DR16 (eg. \citealt{Rojas2019}). We encourage the reader to check Section 7. of \citet{Guiglion2024} for more discussion on this figure. By investigating the [$\alpha$/M] vs. [M/H] pattern over a large range of galactic R and Z, we show that CNN is able to recover the main abundance trends in the Milky Way over a large galactic volume, even for low S/N ratio RVS data. This first detection of the bimodality in the RVS data using a CNN is a step forward in the scientific output of the \emph{Gaia} mission. We refer the reader to \citet{Nepal2024a} and \citet{Nepal2024b} for science exploitation of the RVS/CNN catalog of \citet{Guiglion2024} focused on the old MW thin disc and the bar, respectively). 

\section{Prospects}

The next \emph{Gaia} data release will consist of 66 months of data (expected by the end of 2025) and will include all epoch and transit data for all sources (i.e. low S/N RVS data). CNN for Gaia-RVS represents a step forward in the analysis of such a dataset.
For the next studies and generation of surveys, the training sample should be built in a proactive way, that is, by selecting targets to be observed instead of simply using an existing set of reference stars. In this way, the biases inherent to any training sample will be limited. For instance, focus should be on the tail of metal-poor stars as well as bright stars, local stars, and M giants. Such a challenge will have to be faced by 4MOST, which aims at using ML tools for stellar parameterisation. On that topic, we believe that the experience gained here with the analysis of spectra with limited resolution and spectral coverage will be crucial in the development of a CNN method for future surveys, such as the 4MOST
Milky Way Disc and Bulge Low-Resolution Survey (4MIDABLE-LR; \citealt{Chiappini2019}) or the Wide-field Spectroscopic Telescope (WST, \citealt{Mainieri2024}).

% ------------------------------------------------------------

\acknowledgements
G. G. sincerely thanks the SOC and LOC of the IAUS395 for such an amazing conference. G.G. acknowledges support by Deutsche Forschungs-gemeinschaft (DFG,
German Research Foundation) – project-IDs: eBer-22-59652 (GU 2240/1-1
"Galactic Archaeology with Convolutional Neural-Networks: Realising the
potential of Gaia and 4MOST"). This project has received funding from
the European Research Council (ERC) under the European Union’s Horizon
2020 research and innovation programme (Grant agreement No. 949173).
GG. also thanks the support from the International Astronomical Union grant for the IAUS395 2024. 

% \bibliographystyle{apalike}
% \bibliography{references.bib}

\begin{thebibliography}{}
\bibitem[Abdurro'uf et al. (2021)]{Abdurrouf2022} Abdurro’uf, Accetta, K., Aerts, C., et al. 2022, ApJS, 259, 35.
\bibitem[Ambrosch et al. (2023)]{Ambrosch2023} Ambrosch, M., Guiglion, G., et al. 2023, \aap, 672, A46.
\bibitem[Chiappini et al. (2019)]{Chiappini2019} Chiappini, C., Minchev, I., et al. 2019, The Messenger, 175, 30.
\bibitem[De Angeli et al. (2023)]{DeAngeli2023} De Angeli, F., Weiler, M., et al. 2023, \aap, 674, A2.
\bibitem[de Jong et al. (2019)]{deJong2019} de Jong, R. S., Agertz, O., et al. 2019, The Messenger, 175, 3
\bibitem[Fabbro et al. (2018)]{Fabbro2018} Fabbro, S., Venn, K. A., et al. 2018, MNRAS, 475, 2978.
\bibitem[Gaia Collaboration (2023)]{GaiaCollaboration2023} Gaia Collaboration (Vallenari, A., et al.) 2023, \aap, 674, A1.
\bibitem[Gilmore et al. (2022)]{Gilmore2022} Gilmore, G., Randich, S., et al. 2022, \aap, 666, A120.
\bibitem[Guiglion et al. (2020)]{Guiglion2020} Guiglion, G., Matijevi$\mathrm{\check{c}}$, G., et al. 2024, \aap, 644, A168.
\bibitem[Guiglion et al. (2024)]{Guiglion2024} Guiglion, G., Nepal, S., et al. 2024, \aap, 682, A9.
\bibitem[Leung \& Bovy (2019)]{Leung2019} Leung, H. W., \& Bovy, J. 2019, MNRAS, 483, 3.
\bibitem[Lindegren et al. (2021)]{Lindegren2021} Lindegren, L., Bastian, U., et al. 2021, \aap, 649, A4.
\bibitem[Mainieri et al. (2024)]{Mainieri2024} Mainieri, V., Anderson, R., et al. 2024, eprint arXiv:2403.05398.
\bibitem[Majewski et al. (2017)]{Majewski2017}  Majewski, S. R., Schiavon, R. P., et al. 2017, AJ, 154, 94.
\bibitem[Nepal et al. (2023)]{Nepal2023} Nepal, S., Guiglion, G., et al. 2023, \aap, 671, A61.
\bibitem[Nepal et al. (2024a)]{Nepal2024a} Nepal, S., Chiappini, C., et al. 2024, \aap, 681, L8.
\bibitem[Nepal et al. (2024b)]{Nepal2024b} Nepal, S., Chiappini, C., et al. 2024, \aap, 688, A167.
\bibitem[Matteucci (2021)]{Matteucci2021} Matteucci, F. 2021, A\&ARv, 29, 5.
\bibitem[Rojas-Arriagada et al. (2019)]{Rojas2019} Rojas-Arriagada, A., Zoccali, M., et al. 2019, \aap, 626, A16.

\end{thebibliography}

\end{document}